# Form of spinning liquids in diverse geometries

Paul Menker, and Andrzej Herczyński

# Form of spinning liquids in diverse geometries

Paul Menker and Andrzej Herczyński
*Department of Physics, Boston College, Chestnut Hill, Massachusetts 02467*

A series of experiments for steady state rotation of water in vessels of various geometries is presented. The experiments focus on the geometrical characteristics of the rotating liquids and the change in their surface topology, from that akin to a sphere to that of a torus (i.e., from genus 0 to 1), for sufficiently large angular speeds. Cylindrical, planar rectangular, cubic, spherical, and conical containers are considered. The cone is an exception as some liquid always remains in its apex, no matter how fast the spin. It is shown also that for any amount of liquid within, there exists a critical angular speed above which the liquid can no longer be confined and is therefore expelled from the cone spontaneously breaking the symmetry. This instability is investigated experimentally. © *2020 American Association of Physics Teachers.*
https://doi.org/10.1119/10.0001178

## I. INTRODUCTION

The literature on the behavior of a liquid in rotating containers is extensive. The earliest documented experiment was probably Newton's investigation of a spinning bucket filled with water. Newton recognized that once the pail and the water "moved as one body," the surface of the water takes the shape of a paraboloid of revolution, and he used this observation in his disputes with Leibnitz to argue for the existence of absolute (inertial) reference frames.[1]

A number of straightforward experimental techniques have been devised to demonstrate the parabolic surface profile in the simplest case, when a liquid spins uniformly in a cylindrical vessel.[2–7] The accuracy of these methods does not typically permit discerning capillary effects near the vessel's walls. The effect of the surface tension has been analyzed in a number of papers[8–10] showing that the free surface is flatter near the rim than for a paraboloid. The evolution of the profile of the interface between two immiscible liquids, oil and water, during a spin-up in a rotating cylinder has been studied theoretically and experimentally by Yan *et al.*[11]

Because of their parabolic profiles, spinning liquid mirrors held inside cylindrical containers are important in astronomy and for optical devices.[12,13] The most convenient liquid is mercury (or an alloy of gallium) and the usual application is for the primary mirror in reflecting telescopes. Hickson[14] provided a review of the state-of the-art technology for such telescopes. Extensions to other geometries in the literature are mostly limited to spherical vessels and applications to fuel tanks (e.g., Rabitti and Maas[15]).

The present contribution takes up the steady-state uniform rotation of a liquid and encompasses vessels of five different shapes—planar rectangular, cubic, cylindrical, spherical, and conical—and effects not yet explored in the literature. The experiments are focused on the shape of the spinning liquid for supercritical angular speeds, when the apex of the free surface paraboloid is below the lowest point of the container. The dry region, which develops in this case at the base of the container, is investigated here.

The sphere and the cone represent special cases since they lack flat bases. In the sphere, the way a paraboloidal free liquid surface is accommodated depends critically on whether the upper hemisphere is wetted or not. The conic container is unique among the vessels considered here. Because of its singularity at the apex, a dry region does not appear in a cone no matter how fast the rotation. Furthermore, for any amount of liquid within, there exists the maximum angular speed compatible with the steady state rotation; when that speed is exceeded, an instability develops. The critical angular speed is measured and the instability is described qualitatively.

The plan of presentation is as follows: in Sec. II, the approach used is formulated and the notation is introduced. In Sec. III, critical angular speeds and other parameters for all geometries are obtained. Section IV presents all experimental results, which are discussed in concluding Sec. V.

## II. PROBLEM FORMULATION

Consider a liquid in a container rotating at angular speed $\omega$ and assume that a steady state has been reached so that the liquid rotates as a solid body, with the uniform angular speed of the vessel. Ignoring surface tension, elementary considerations[3,4,6] of the forces acting on a fluid element at the liquid's free surface show that the local slope, $dz/dr$, is given by the ratio of the centripetal force to the weight of the fluid element, and so the surface will take the shape of the paraboloid of rotation given in cylindrical coordinates by

$$z(r) = \alpha r^2 + d, \quad (1)$$

where the rotation parameter $\alpha$ (steepness of the paraboloid) is defined as

$$\alpha = \frac{\omega^2}{2g}. \quad (2)$$

Choosing $z = 0$ as the position of the container's lowest point, the constant of integration $d$ in Eq. (1) represents the depth of the liquid along the axis of rotation, which depends on the shape of the container, the volume of the liquid, and $\omega$.

Let $h$ denotes the initial depth of the liquid along the z-axis when at rest. At the critical angular speed $\omega_1$, the depth at the center $d = 0$. It follows that $d > 0$ corresponds to the subcritical case, when $\omega < \omega_1$, and $d < 0$ to the supercritical case, when $\omega > \omega_1$.

The main focus here is on the critical angular speed $\omega_1$ and the depth $d(\omega)$ for subcritical, and the radius of the dry patch $\delta(\omega)$ for the supercritical cases. For the cone, the aim is to measure the maximum angular speed $\omega_0$ possible for a given volume of liquid. Since only steady-state motion at uniform angular speed is considered, the problem is geometrical, that of volume conservation under the free surface paraboloid.

To keep the analysis physically transparent, it is presented in dimensional form; however, experimental results (except those for the cone) are plotted in dimensionless form. Table I lists the principal parameters of the problem and their dimensionless counterparts based on $R$, either the radius (cylinder and sphere) or its equivalent, the half-length (rectangle and cube).

## III. ALGEBRAIC SOLUTIONS

### A. Cylindrical geometry

Let a stationary cylinder of radius $R$ be filled to the level $h$ with a liquid so that the volume occupied is $V = \pi R^2 h$. When the cylinder is rotating and the liquid is in steady state in the subcritical case, with the depth along the axis of symmetry $z$ denoted as $d$, the volume contained can be expressed as $V = \alpha \pi R^4/2 + \pi R^2 d$. Equating these two expressions and writing out the rotational parameter $\alpha$ explicitly yield

$$d = h - \frac{\omega^2 R^2}{4g}. \tag{3}$$

For the supercritical case, with a dry patch of radius $\delta$ at the bottom, $d = -\alpha\delta^2$ and the free surface is given by $z = \alpha(r^2 - \delta^2)$. The volume of the liquid under this paraboloid, $V = \alpha\pi R^4/2 - \alpha\pi\delta^4/2 - \alpha\pi\delta^2(R^2 - \delta^2)$, must be equal to the initial volume, so that

$$\delta^4 - 2R^2\delta^2 + R^2\left(R^2 - \frac{2h}{\alpha}\right) = 0, \tag{4}$$

whose physically meaningful solution (positive and smaller than $R$) is

$$\delta = R\sqrt{1 - \frac{2\sqrt{hg}}{R\omega}}. \tag{5}$$

Now setting either $d = 0$ in Eq. (3), or $\delta = 0$ in Eqs. (4) or (5), gives the critical angular speed as

$$\omega_1 = \frac{2}{R}\sqrt{hg}. \tag{6}$$

### B. Planar geometry

Consider a rectangular container of side $2R$ and width $w$ (and sufficient height) so the volume of liquid within is $V = 2Rhw$. For a given subcritical angular speed $\omega$, the free surface is defined by $z = \alpha(x^2 + y^2) + d$. Integrating this expression over the range of $x$ and $y$ leads to the volume conservation condition

Table I. Dimensionless parameters.

|  | Dimensional | Dimensionless |
| --- | --- | --- |
| Angular speed | $\omega$ | $\Omega = \omega\sqrt{R/g}$ |
| Static depth | $h$ | $H = h/R$ |
| Dynamic depth | $d$ | $D = d/R$ |
| Dry radius | $\delta$ | $\Delta = \delta/R$ |

$$d = h - \frac{\omega^2 R^2}{6g}\left(1 + \frac{w^2}{4R^2}\right). \tag{7}$$

For the planar case, the width is assumed small compared to the length, $w \ll 2R$, so

$$d = h - \frac{\omega^2 R^2}{6g}. \tag{8}$$

In the experiments described here (see Sec. IV), the error introduced by using Eq. (8) instead of Eq. (7) for the planar-rectangular container is at most 0.4%, smaller than the experimental uncertainty in the corresponding measurements.

In the supercritical case, with $d = -\alpha\delta^2$, the free surface is given by $z = \alpha(x^2 + y^2 - \delta^2)$, and the appropriate integration (for the planar case) yields

$$\delta^3 - \frac{3}{2}R\delta^2 + \frac{1}{2}R\left(R^2 - \frac{3h}{\alpha}\right) = 0. \tag{9}$$

Setting $d = 0$ in Eq. (7) gives the critical angular speed

$$\omega_1 = \frac{1}{R}\sqrt{\frac{24R^2hg}{4R^2 + w^2}}, \tag{10}$$

or, in the planar approximation, $w = 0$, obtainable also from Eqs. (8) and (9)

$$\omega_1 = \frac{1}{R}\sqrt{6hg}. \tag{11}$$

The discriminant of Eq. (9) is positive so there are three real roots. The sole physically meaningful root, satisfying $0 \leq \delta \leq R$, can be written in Viète form

$$\delta = R\left(\frac{1}{2} - \cos\frac{1}{3}\left(\pi + \cos^{-1}\left(\frac{12hg}{R^2\omega^2} - 1\right)\right)\right), \tag{12}$$

where the inverse cosine is assumed to be nonnegative. Equation (11) can now be obtained alternatively from Eq. (12) by setting $\delta = 0$.

### C. Cubic geometry

The critical angular speed $\omega_1$ and the subcritical depth $d$ for the cube of side $2R$ can be obtained immediately from the corresponding equations for a planar rectangular container, Eqs. (10) and (7), by setting $w = 2R$. Thus

$$\omega_1 = \frac{1}{R}\sqrt{3hg}, \tag{13}$$

$$d = h - \frac{\omega^2 R^2}{3g}. \tag{14}$$

In the supercritical case, the free surface is again given by $z = \alpha(x^2 + y^2 - \delta^2)$, but the integration (now in both $x$ and $y$) gives a quadratic equation for $\delta^2$

$$\pi\delta^4 - 8R^2\delta^2 + \frac{16}{3}R^4 - \frac{8R^2h}{\alpha} = 0. \tag{15}$$

The solution of Eq. (15) consistent with Eq. (13) is

$$\delta = \frac{2R}{\sqrt{\pi}} \sqrt{1 - \sqrt{1 - \pi\left(\frac{1}{3} - \frac{gh}{R^2\omega^2}\right)}}. \qquad (16)$$

In the cube, one can also distinguish the second critical condition defined by $\delta = R$, and the corresponding second critical angular speed obtained from Eq. (16)

$$\omega_2 = \frac{4}{R}\sqrt{\frac{3hg}{3\pi - 8}}. \qquad (17)$$

For $\omega > \omega_2$, the dry patch at the bottom of the cube is no longer circular. As $\omega$ increases, the dry region gradually evolves becoming squarer and is confined by four straight edges of length $l = 2\sqrt{\delta^2 - R^2}$ each, approaching (but never reaching) $2R$. The parameter $\delta$ is in this case a solution of the transcendental equation

$$\delta^4\left(\frac{\pi}{2} - 2\cos^{-1}\left(\frac{R}{\delta}\right)\right) + 4R^2\delta^2\left(\sqrt{\frac{\delta^2}{R^2} - 1} - 1\right)$$
$$+ 2R^4\left(\frac{4}{3} - \frac{1}{3}\left(\frac{\delta^2}{R^2} - 1\right)^{3/2} - \sqrt{\frac{\delta^2}{R^2} - 1}\right) - \frac{8R^2gh}{\omega^2} = 0. \qquad (18)$$

Setting $\delta = R$ in Eq. (18) recovers Eq. (17), whereas $\delta \to \sqrt{2}R$ in the limit $\omega \to \infty$. In this limit, "squaring of the circle" would be complete.

### D. Spherical geometry

Let a stationary sphere of radius $R$ be filled to the level $h$ with a liquid so that the volume occupied is $V = \pi h^2(R - h/3)$. When the system is rotating, it is convenient to express the relevant parameters in terms of the wetting radius $r_0$. The height of the free surface paraboloid is then $h_0 = \alpha r_0^2$. For the subcritical case, the depth of the spinning liquid along the axis of rotation is

$$d = R - h_0 \pm \sqrt{R^2 - r_0^2}. \qquad (19)$$

The negative sign in (19) corresponds to the case when the upper hemisphere is dry and the positive sign to the case when it is wet. Setting $d = 0$ and $r_0^2 = h_0/\alpha = 2gh_0/\omega^2$ in Eq. (19) and rearranging the resulting expression provides the critical angular speed for the sphere

$$\omega_1 = \sqrt{\frac{2g}{2R - h_0}} = \sqrt{\frac{2g}{2R - (2h^2(3R - h))^{1/3}}}. \qquad (20)$$

In the second part of Eq. (20), $h_0$ is expressed in terms of the initial depth $h$.

Equation (19) reduces the problem of finding $d$ to that of finding $r_0$. Comparing expressions for volumes at rest and when the system is rotating yields after some algebra

$$4R^3 - 2h^2(3R - h) \pm 2(2R^2 + r_0^2)\sqrt{R^2 - r_0^2} - 3\alpha r_0^4 = 0. \qquad (21)$$

Equation (21) is effectively the fourth order polynomial for $r_0^2$ and will be solved numerically. When the wetting circle has the radius $r_0 = R$, Eq. (21) gives

$$\omega_2 = 2\sqrt{\frac{g}{3R}\left(2 - \frac{h^2}{R^2}\left(3 - \frac{h}{R}\right)\right)}. \qquad (22)$$

For $\omega < \omega_2$, the upper hemisphere is dry; for $\omega > \omega_2$, the upper hemisphere is wet.

In the supercritical case, Eq. (19) still applies but now the depth $d$ is negative. The condition of volume conservation leads to the third order polynomial equation for variable $y = \sqrt{R^2 - \delta^2}$ in the form

$$4\alpha^3 y^3 - 6\alpha^2 y^2 + 3\alpha y - \alpha^3 h^2(3R - h) - 1/2 = 0. \qquad (23)$$

It is noteworthy that for the supercritical case a single equation suffices to find $\delta$, valid whether there is liquid above the equator or not. The discriminant of Eq. (23) is negative so that only one real solution for $y$ exists, and consequently, $\delta = \sqrt{R^2 - y^2}$ is uniquely determined. With $\delta = 0$, or $y = R$, Eq. (20) is the solution of Eq. (23).

### E. Conic geometry

There is no characteristic length-scale for a cone and therefore no natural normalization of the equations for liquid moving within. Here, a right cone was considered, (the actual apex angle of the cone used in the experiments was 89°, see Sec. IV). Since for a right cone $z = r$, the wetting radius is given by

$$r = \alpha r^2 + d. \qquad (24)$$

Solving this quadratic equation gives

$$r = \frac{1}{2\alpha} - \sqrt{\frac{1}{4\alpha^2} - \frac{d}{\alpha}}, \qquad (25)$$

where the second, larger solution was discarded since it does not represent the liquid contact line. It follows from Eq. (25) that $d = 0$ if and only if $r = 0$; liquid cannot completely escape from the cone's apex at steady-state. From Eq. (25) it follows also that, for any angular speed $\omega$, there exists a maximum allowable wetting radius $r_0 = 1/2\alpha$ corresponding to the maximum allowable depth $d_0 = 1/4\alpha$. This conclusion can be reformulated as follows. For any volume of liquid in the cone, there exists the critical $\omega_0$, such that for $\omega > \omega_0$ the liquid can no longer be contained and, consequently, will become unstable and spill. This can be seen directly by noting that the volume "available" in the cone under the free surface of wetting radius $r$ is

$$V = \pi r^3 \left(\frac{1}{3} - \frac{\alpha r}{2}\right). \qquad (26)$$

The maximum available volume is therefore obtained by setting $r = 1/2\alpha$ in Eq. (26) or

$$V_{\max} = \frac{\pi}{96\alpha^3} = \frac{1}{3}\pi h^3, \qquad (27)$$

where $h$ is the initial depth of the liquid in the cone at rest. Equation (27) furnishes the critical angular speed, above which the liquid will become unstable for a given $h$

$$\omega_0 = 2^{-1/3}\sqrt{\frac{g}{h}}. \quad (28)$$

For angular speeds below $\omega_0$, the depth $d$ is obtained from the corresponding radius $r$ and using Eq. (24). Setting $V = \pi h^3/3$ in Eq. (26)

$$r^4 - \frac{2}{3\alpha}r^3 + \frac{2}{3\alpha}h^3 = 0. \quad (29)$$

The discriminant for Eq. (29) reveals that there is one (double) real root if $32h^3 - 1/\alpha^3 = 0$ and there are two real roots (of which the smaller is the one physically meaningful here) when $32h^3 - 1/\alpha^3 < 0$; otherwise, there are no real solutions. This confirms the existence of the maximum allowable angular speed for steady state given by Eq. (28). While solutions to Eq. (29) can be written out explicitly, they are cumbersome and are omitted here.

Equation (28) can be easily generalized to the case of the cone of arbitrary apex angle $\beta$. The result is

$$\omega_0 = 2^{-1/3}\sqrt{\frac{g}{h}\frac{1}{\tan(\beta/2)}}. \quad (30)$$

This suggests a non-dimensional parameter for the spinning cone of apex angle $\beta$

$$\mathrm{Cn} = \frac{\omega^2 h}{g}\tan^2(\beta/2). \quad (31)$$

When $\mathrm{Cn} > \mathrm{Cn}_0 = 2^{-2/3} \approx 0.63$, the liquid becomes unstable.

## IV. EXPERIMENTAL PROCEDURE

The experimental setup utilized a commercially available PASCO rotating platform outfitted with an electric motor. Attached to the platform was an $L$-shaped trigger for a photo-gate used to measure the period of rotation. Data, in the form of period vs time, were collected automatically. Two versions of the general setup are shown in Figs. 1 and 2.

Each of the five acrylic containers was attached directly to the rotating platform. The planar-rectangular container is available from PASCO. The other four vessels were either acquired from commercial vendors (cube, cylinder, and the sphere) or manufactured (the cone). To each a base was glued, which could be fastened with screws to the rotating platform. Dimensions of the five containers are given in Table II.

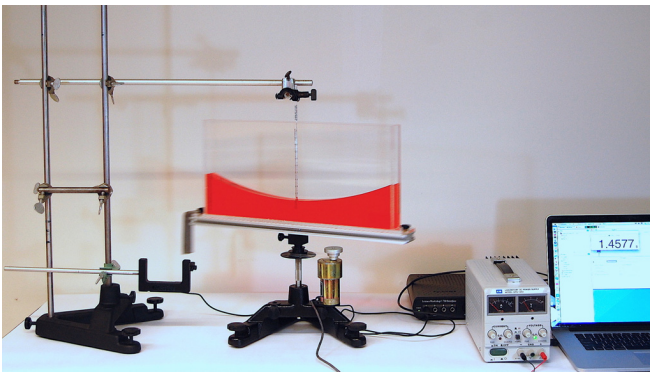

Fig. 1. Experimental setup showing a planar rectangular container attached to the rotating platform and spinning with the subcritical speed (period $T = 1.4557\,\mathrm{s}$) and the pipette positioned to measure depth $d$ at the axis of rotation. This method was actually used for the cylinder, the cube, the sphere, and the cone.

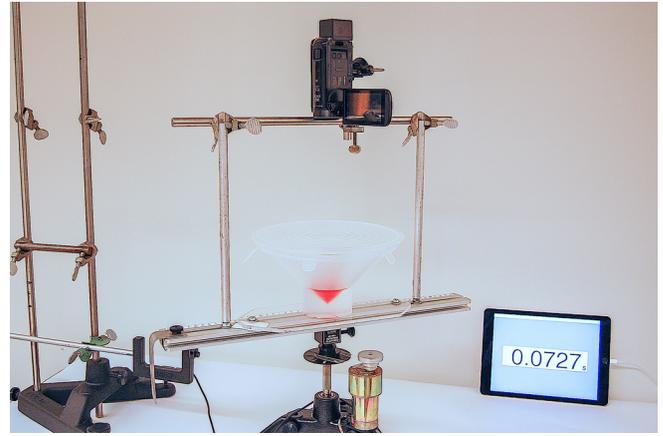

Fig. 2. Experimental setup used for filming the instability in the cone, with the camera positioned above the vessel and attached to the rotating platform. Here, the stationary depth of water was $h = 3$ cm. The tablet was used for displaying the period $T$.

The initial depth of water in each container was obtained from the mass of water poured inside. De-ionized water was used and weighed on a balance with the accuracy of the tenth of a gram. For every 1 kg of water, approximately, ten drops of red food coloring were added and also 5 g of Kodak Photo-Flo 200 solution to reduce the surface tension.

For subcritical angular speeds, two different measuring techniques were used to obtain the depth at the centerline. For the planar rectangular container, side view photographs were taken with a camera (Nikon D40 with AF-S DX Zoom-Nikkor lens) attached to a stationary frame. The photos were then processed using the *Photos* application. Distances were measured in pixels and then converted to centimeters based on known dimensions of the container. At each depth and angular speed, multiple photos were taken and the measurements were averaged for each data point.

For the cylinder, cube, sphere, and cone, a different method was used. A very narrow (diameter 3 mm) graduated pipette, attached to a fixed frame, was inserted along the centerline and used as a ruler. The pipette's gradation in ml was then converted to centimeters (0.1 ml corresponds to 1.443 cm). This technique can only provide relative depths, but one can compare the reading at rest with the known depth to obtain the absolute measurements when the system is rotating.

In measurements of the critical speeds, it was found that the best method is to approach the critical condition from above. The gradual closing of the dry spot was then observed using a bright LED light from behind the container. The dry spot reflected that light while the liquid did not so the critical angular speed could be accurately determined. For each container and water depth, four measurements at the critical condition were averaged: two with the clockwise and two with the counterclockwise spin.

Table II. The inside dimensions of containers in centimeters.

|  | Rectangle | Cylinder | Cube | Sphere | Cone |
|---|---|---|---|---|---|
| Radius (cm) | 21.76 | 9.91 | 12.44 | 12.65 | N/A |
| Width (cm) | 1.44 | N/A | N/A | N/A | N/A |
| Height (cm) | 20.35 | 24.47 | 25.09 | N/A | 10.43 |
| Apex angle (°) | N/A | N/A | N/A | N/A | 89.0 |

For supercritical measurements, photographs of dry regions were taken from above the containers. The horizontal diameter of the dry spot was measured in pixels and then converted to centimeters as before.

In the cylinder, the vertical position of the surface at the center for the subcritical case was measured from below through the water. The dimensionless version of Eq. (6) for the critical angular speed in the cylinder is simply $\Omega_1 = 2\sqrt{H}$, and Eqs. (3) and (5) are rendered in non-dimensional form as $D = H - \Omega^2/4$, $\Delta = \sqrt{1 - 2\sqrt{H}/\Omega}$. Critical measurements for the cylinder are shown in Fig. 3, whereas subcritical and supercritical results are compared with the above two expressions in Fig. 4.

For the planar rectangular container, the narrow width (1.44 cm) made the critical angular speed measurements harder. To compensate, four measurements were taken starting below the critical condition and four starting above. Useful photographs could only be taken when the container's plane was parallel to the camera's lens so all other photographs were discarded. The critical angular speed in this case, Eq. (11), can be written in dimensionless form as $\Omega_1 = \sqrt{6H}$, and the dimensionless form of Eqs. (8) and (12) is readily obtained. Critical measurements for the rectangle are shown in Fig. 3, and the subcritical and supercritical results are shown in Fig. 5.

For the cube, the first and the second critical angular speeds were measured using a similar approach to that for the cylinder. Data were also taken above the second critical speed in the cube, whereby the radius was measured indirectly, from the lengths of the four straight boundaries of the dry region. Three images were obtained for each angular speed and the four "corner" lengths were averaged for each data point. The dimensionless versions of Eqs. (13) and (17) are $\Omega_1 = \sqrt{3H}$ and $\Omega_2 = 4\sqrt{3H/(3\pi - 8)}$, respectively, and the dimensionless forms of Eqs. (14) and (16) are similarly obtained. Experimental values for the two critical angular speeds are shown in Fig. 3. Figure 6 displays subcritical and supercritical data against theoretical curves, the latter both for $\Omega_1 \leq \Omega \leq \Omega_2$ and $\Omega_2 \leq \Omega$.

For the sphere, most data were taken through the water, but near the critical angular speed the readings were taken above the water surface. The dimensionless critical angular speed corresponding to Eq. (20) is $\Omega_1 = \sqrt{2/(2 - (2H^2(3 - H))^{1/3})}$. Figure 7 shows experimental values of $\Omega_1$ at various fill levels $H$. In Fig. 8, measurements for subcritical and supercritical cases are compared with the theoretical curves. The formula correcting for parallax error in the supercritical data is derived in the Appendix.

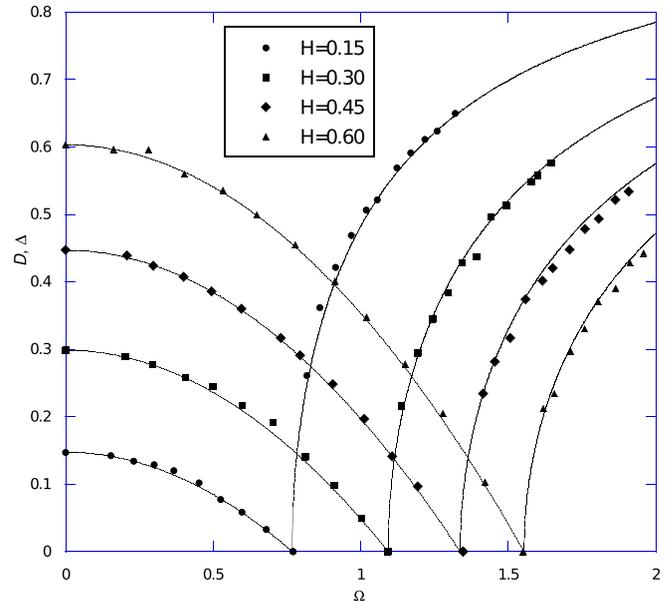

Fig. 4. Subcritical depth $D$ and the radius of the dry region $\Delta$ for the cylinder at four initial fill levels $H$ as a function of $\Omega$. Continuous lines are plots of Eqs. (3) and (5).

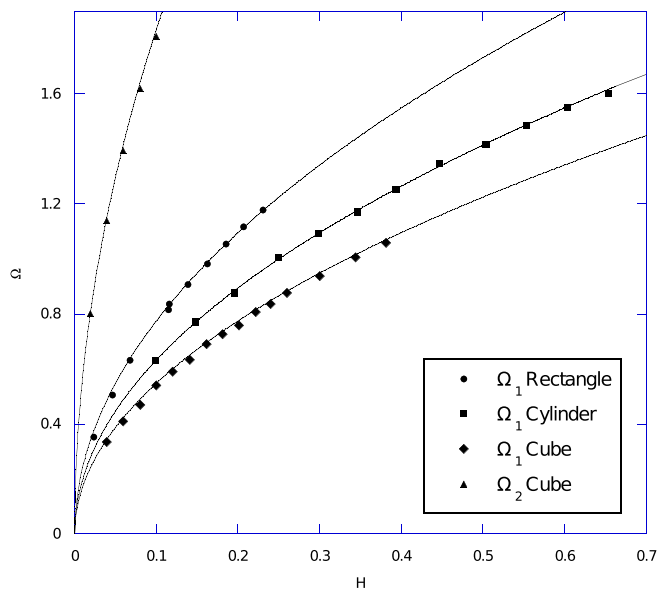

Fig. 3. Critical angular speed $\Omega_1$ for the cylinder and the rectangle, and the two critical speeds $\Omega_1$ and $\Omega_2$ for the cube as a function of the fill level $H$. Continuous lines correspond to Eqs. (6), (11), (13), and (17), respectively.

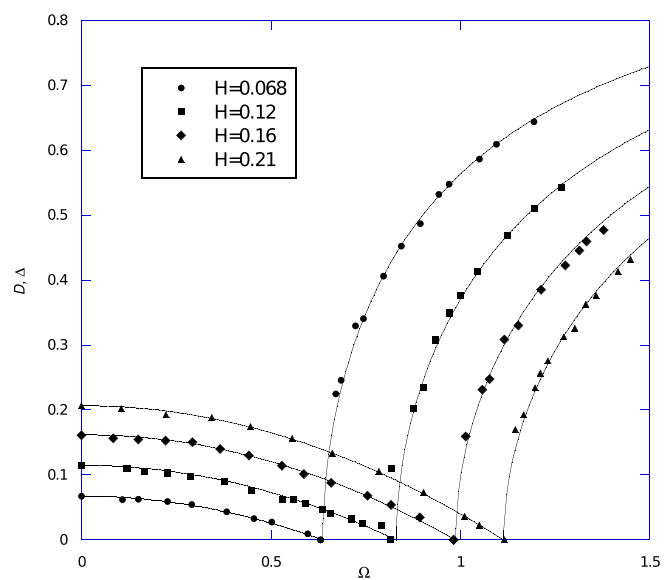

Fig. 5. Subcritical depth $D$ and the radius of the dry region $\Delta$ for the rectangular container at four initial fill levels $H$ as a function of $\Omega$. Continuous lines are obtained from Eqs. (8) and (12).

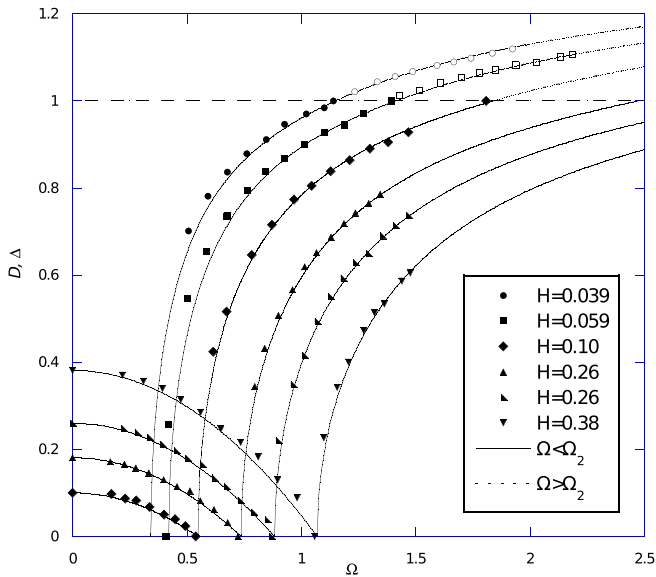

Fig. 6. Subcritical depth $D$ and the radius of the dry region $\Delta$ for the cube at four initial fill levels $H$ as a function of $\Omega$. Open symbols represent data for $\Omega > \Omega_1$. Continuous lines are obtained from Eqs. (14) and (16); dashed lines are numerical solutions of Eq. (18).

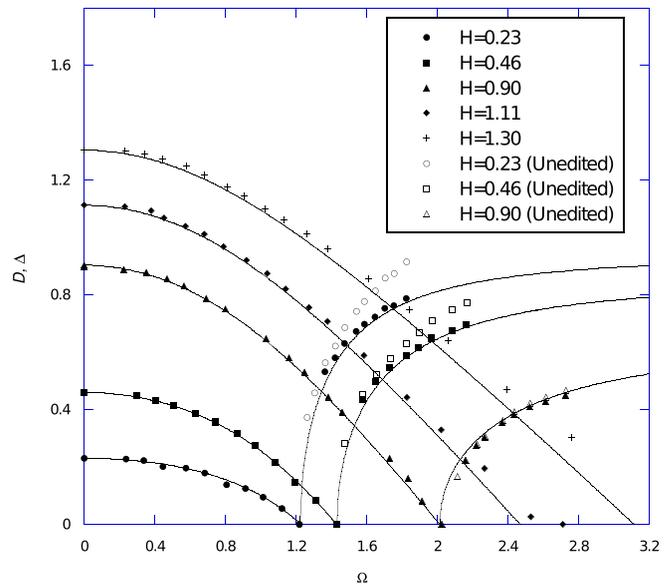

Fig. 8. Subcritical depth $D$ and the radius of the dry region $\Delta$ for the spherical container at four initial fill levels $H$ as a function of $\Omega$. Continuous lines are obtained from Eq. (19) with the numerical solutions of Eq. (21) and from Eq. (23).

Since there is always some water left in the apex of the cone in steady state, experiments conducted focused on the depth of water along the axis and the instability that leads to liquid spilling out. Figure 9 shows the depth as a function of angular speed obtained from the solutions to Eqs. (29) and (24). The measurements were stopped before the critical angular speed, given in Eq. (28), was reached.

A separate series of experiments was performed to elucidate the instability leading to an expulsion of water from the cone. In principle, this will occur for cones of any apex angle at sufficiently large angular speeds regardless of the cone's size. In a real, imperfect apparatus, instability develops at first gradually, in the form of a liquid tongue, which starts as a small, local protrusion on one side of the water contact line, breaking the circular symmetry. This tongue grows gradually with the increasing rate of rotation and at the critical angular speed, given by Eq. (28), suddenly and rapidly spirals up and around the cone's wall until some amount of water spills over the edge.

In the measurements, the onset of this spiral was taken to indicate that the critical condition was reached, and the corresponding angular speed was recorded. Critical angular speeds obtained this way are plotted in Fig. 10. Videos of the process from above with the camera attached to the cone were taken at various fill-levels. It was noted that the spill was more rapid for smaller volumes of water in the cone. A video of the process for the initial depth of $h = 3$ cm is available online as the supplementary material.[16]

## V. DISCUSSION AND CONCLUSIONS

While the existing literature on steady state rotation of liquids is largely limited to cylindrical containers and subcritical angular speeds,[2–14] the experiments described here explore, in four different geometries, how a dry region near the container's lowest point opens up above the critical angular speed.

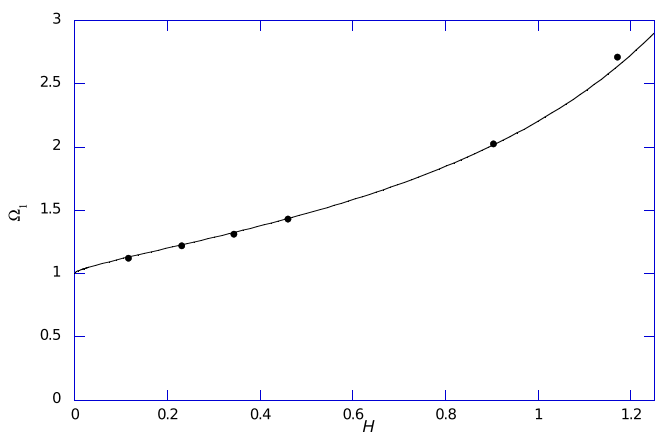

Fig. 7. Critical angular speed $\Omega_1$ for the sphere as a function of the fill level $H$. The continuous line is obtained from Eq. (20).

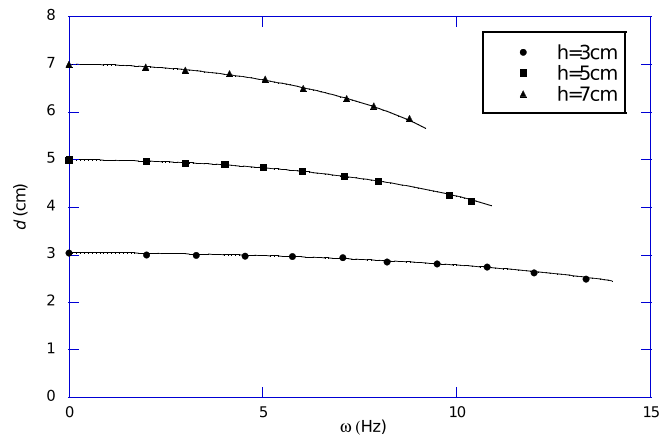

Fig. 9. Subcritical depth $d$ for the cone for three different initial fill levels $h$ as a function of angular speed $\omega$. Continuous lines are obtained from Eqs. (29) and (24).

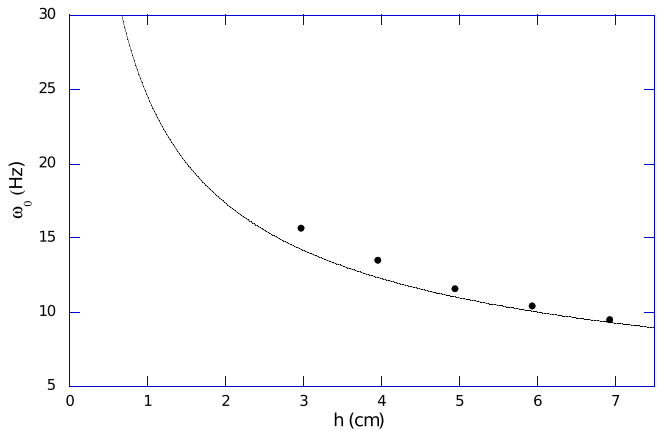

Fig. 10. Critical angular speed $\omega_0$ for the cone as a function of $h$ plotted against Eq. (28). For details on how these experimental data were obtained, see the last paragraph of Sec. IV.

For the containers with flat bases—planar, cylindrical, and cubic—this topological transition is governed by algebraic equations of the third or fourth order. Experimental values of the critical angular speeds, and the characteristics of the liquid surfaces below and above criticality, agree well with the analytical predictions indicating that, after a short accommodation time, the liquid does spin as a rigid body with the surface tension playing a negligible role. The similarity of the plots for the subcritical and supercritical behavior in different geometries (Figs. 4–6) belies their substantially different mathematical form, as seen for example by comparing Eqs. (5), (12), (16), and the transcendental Eq. (18).

The dependence of the critical speed on the initial fill level in the sphere departs from the behavior in containers with flat bottoms. All critical curves in Fig. 3 start at the origin and have negative curvatures and decreasing slopes. For the sphere, the corresponding curve in Fig. 7 has an altogether different character and two remarkable, subtle features. First, implausibly on the first glance, the critical angular speed does not tend to zero as the fill level $H$ goes to zero. This can be understood as follows. As the volume decreases, the free surface curvature, $2\alpha$, must approach that of the sphere at the lowest point, $1/R$, rather than zero. When the two curvatures are the same, at vanishing volume, $2\alpha = 1/R$ or $\omega^2 = g/R$, which defines the vertical axis intercept in Fig. 7. Any finite volume of water implies that $\omega^2 > g/R$ or $\Omega^2 > 1$ at $D = 0$; otherwise, there would be no "space" for the liquid under the free surface.

The second notable feature is the change in curvature of the plot in Fig. 7 as given by Eq. (20). For sufficiently small volumes, the curve is concave down, as for the containers with vertical walls. However, the curvature becomes positive for larger fill levels, and there is a singularity at $h \to 2R$ corresponding to a filled sphere. It can be seen that the inflection point corresponds to the critical angular speed at which the upper hemisphere just becomes wet. Indeed, this will occur when $\omega_1 = \omega_2$, that is when $h_0 = R$ or $H_0 = 1$, as one would expect (see Eqs. (20) and (22)). Using the expression for $H_0$ in terms of $H$, this condition corresponds to $H \approx 0.45$, in agreement with the apparent position of the inflection point in Fig. 7. The data for the sphere had to be corrected for the parallax error. Figure 8 shows both the original data ("unedited") and the corrected values, using Eq. (A3) in the Appendix, which agree well with the theoretical curves.

In general, the accuracy of all measurements decreased with increased initial depth of water in the container, which is especially evident in the subcritical data for the sphere in Fig. 8. This is because the critical speed increases with the volume making reading scales harder, particularly in this geometry.

For cylindrical, planar rectangular, and cubic containers, the geometrical parameters $d$ and $\delta$ of the paraboidal free-surfaces can be used to obtain the minimum heights $s$ of these vessels required to avoid spills. For the simplest case of the cylinder, for example, Eq. (1) gives $s = h + \omega^2 R^2/4g$ for subcritical speeds, and (setting $d = -\alpha\delta^2$) $s = \omega R\sqrt{h/g}$ for supercritical speeds, so that $s = 2h$, independent of $R$, for $\omega = \omega_1$. Spill heights for containers of other shapes can be obtained in a similar manner.

In regard to planar rectangular and cylindrical containers, it may be useful to note that they can be both viewed as special cases of an $n$-dimensional cylinder, whose "base" is a hypersphere of dimension $n - 1 (n > 1)$ with $n = 2$ and $n = 3$, respectively. Indeed, Eqs. (3) and (8) are special cases of $d = h - \omega^2 R^2 (n-1)/2g(n+1)$, whereas Eqs. (6) and (11) are special cases of $\omega_1^2 = 2(n+1)hg/(n-1)R^2$.

To conclude, a few comments on the behavior in the cone, the most intriguing case, are in order. The cone is fundamentally different from other containers considered here. It has a singularity at the apex and as a result there is always some liquid left there and the topological transition does not occur. Because its wall is inclined, there exists the maximum angular speed $\omega_0 = \omega_0(h)$ for any initial depth $h$, given by Eq. (28) (or Eq. (30) more generally) above which the liquid can no longer be contained in the cone and therefore must be expelled.

A dimensionless parameter, denoted here $Cn$, has been proposed to characterize the onset of the instability leading to such a spill of the liquid. The manner in which this instability develops must necessarily involve a spontaneous symmetry breaking, a phenomenon that occurs also in a number of other systems involving rotating liquids such as the coiling instability, the Rayleigh–Benard instability, or the Taylor–Couette instability. The process in the cone is described qualitatively in Sec. IV but its quantitative analysis is beyond the scope of the present work.

## ACKNOWLEDGMENTS

The authors would like to thank Krastan Blagoev for suggesting the problem of spinning liquid in a sphere and Patrick Weidman for proposing the cone as another vessel. Thomas Kempa collected preliminary data in rectangular and cylindrical containers when he was an undergraduate student at Boston College. More recently, students Justin Joss and Nick Kearly helped with the experiments. Thanks are also due to Alex Shvonski for his assistance with taking the videos of the instability in the cone. Finally, the authors would like to express gratitude to Krzysztof Kempa for many helpful discussions.

## APPENDIX: PARALLAX CORRECTION

The measurements of the dry region radius in the sphere suffered parallax error, depending on the fill level and angular speed, and reaching up to 12% (see Fig. 8). This error was caused by the fact that the circular wetting line moved closer to the camera's objective as the angular speed increased while its radius was measured in the photographs using a fixed scale below the sphere.

Let $\delta'$ denote the (exaggerated) radius of the dry-wet contact circle as seen by the camera against the fixed horizontal ruler. From the geometry, it is easy to see that

$$\frac{\delta}{\delta'} = 1 - \frac{1}{l}\left(R - \sqrt{R^2 + \delta^2}\right), \quad (A1)$$

where $l$ is the distance from the sphere's lowest point inside to the focus point in the camera. Rewriting Eq. (A1) as a quadratic equation for $\delta$

$$\delta^2\left(\frac{1}{l^2} - \frac{1}{\delta'^2}\right) + \delta\frac{2}{\delta'}\left(\frac{R}{l} - 1\right) - \frac{2R}{l} + 1 = 0. \quad (A2)$$

The physically appropriate solution of Eq. (A2) can be written in dimensionless form as

$$\Delta = \frac{\Delta'}{\Delta'^2 + L^2}\left(L - 1 + \sqrt{1 - \Delta'^2\left(1 - \frac{2}{L}\right)}\right), \quad (A3)$$

where all the lengths were normalized by the radius of the sphere, $\Delta = \delta/R$, $\Delta' = \delta'/R$, and $L = l/R$.

Equation (A3) requires the value of $L$. While the camera was fixed above the spherical container for all measurements, its zoom was adjusted raising the question of whether this affected the effective focal length. A series of experiments were conducted to test this, whereby thin aluminum rods of various known lengths $2w$ ($w < R$) were placed horizontally in the dry sphere and the zoom was adjusted so that the rods spanned the field of view. Let $l/w = \gamma Z$, where $Z$ is the numerical value of the zoom. Measurements at fixed positions of the camera showed that $\gamma$ is in fact a constant ($\gamma = 0.08 \pm 0.01$ here). This indicates that $w \propto 1/Z$, or the effective focal length $l = \gamma w Z$ is independent of the zoom. For the experiments conducted here, $l = 38$ cm was set so that $L = 3.0$.